# What are the most important factors that influence the changes in London Real Estate Prices? How to quantify them?

**Yiyang Gu (yiyang.gu@ucl.ac.uk)**
MSc Smart Cities and Urban Analytics
The Bartlett Centre for Advanced Spatial Analysis

## Abstract

In recent years, real estate industry has captured government and public attention around the world. The factors influencing the prices of real estate are diversified and complex. However, due to the limitations and one-sidedness of their respective views, they did not provide enough theoretical basis for the fluctuation of house price and its influential factors. The purpose of this paper is to build a housing price model to make the scientific and objective analysis of London's real estate market trends from the year 1996 to 2016 and proposes some countermeasures to reasonably control house prices. Specifically, the paper analyzes eight factors which affect the house prices from two aspects: housing supply and demand and find out the factor which is of vital importance to the increase of housing price per square meter. The problem of a high level of multicollinearity between them is solved by using principal components analysis.

## 1 Introduction

In recent years, the development of real estate industry has become an important driving engine of economic growth, but the real estate industry is also suffered criticism (Pyhrrey et al., 2004). The factor that the price of house is unaffordable draws the government and public attention (Case and Shiller, 2003). The factors influencing the prices of real estate are diversified and complex.

Regarding the factors affecting real estate costs, scholars have conducted quite a few researches: Case and Shiller (1990) performed regression analysis and proved that housing price was correlated with population, income, and real estate market profits. Poterba (1991) chose income, building cost and the population as influencing factors, and concluded that the costs could be explained through them. Quigley (1999) emphasized that some economic changes such as population, employment, the economy index could predict the housing price trend. Those all proved that housing price can be modeled and predicted. However, due to the limitations and one-sidedness of their respective views, they did not provide enough theoretical basis for the fluctuation of house price and its influential factors. Moreover, they failed to provide a systematically suitable house price model.

Therefore, this work aims to build a housing price model to make the scientific and objective analysis of London's real estate market trends from the year 1996 to 2016 and proposes some countermeasures to reasonably control house prices. I analyze eight factors which affect the house prices from two aspects: housing supply and demand and find out the factor which is of vital importance to the increase of housing price per square meter. The problem of a high level of multicollinearity between them is solved by using principal components analysis.

Firstly, the available attributes assembled were introduced. The second section explains the multi-regression methodology and principal components analysis, followed by interpretation. The fourth section points out the limitations and conclusion are addressed in the final part.

## 2 Data

Many factors are effecting the real estate costs. Yihong (2016) introduced Real estate investment, Land price, loan interest rates and completed residential area as the variables to describe the supply model, Population, GDP and income as variable indicators to describe the model of demand. Considering the needs of research and data availability, from the above variables, I choose real estate investment, interest rate as my supply factors; population, GVA, and income as demand factors. The variables with sources that apply to the analysis are shown in figure 1.

**Fig.1 Selected variables**

| Variable | Notation | Unit | Source |
|---|---|---|---|
| Changes in House Price (1996-2016) | IY | £ per $m^2$ | ONS and Land Registry |
| Real estate development investment | REI | £ | Greater London Authority（GLA） |
| permanent dwellings started | PDS | in 1,000s | National House-Building Council (NHBC) |
| permanent dwellings completed | PDC | in 1,000s | National House-Building Council (NHBC) |
| Interest Rate | IR | % | Office for National Statistics（ONS） |
| Gross value added | GVA | £ per capita | Office for National Statistics（ONS） |
| Consumer price inflation | CPI | % | Office for National Statistics（ONS） |
| Population per square kilometer | PD | per $km^2$ | Greater London Authority（GLA） |
| Gross disposable household income | GDHI | £ per capita | Office for National Statistics（ONS） |

The reasons why those factors are chosen are following. As for the residence development investment (REI), the more funds the government invests, the larger the building-scale is (Murialdo., 2013). Besides, people usually purchase real estate through mortgage loans. Therefore, interest rate (IR) is bound to be the factor affecting price fluctuations. For the demand factors, the population density (PD) is strongly correlated with the demands of housing purchasing. Gross value added (GVA) is an objective indicator reflecting citizen's payment capacity. Generally speaking, the larger the variable is, the stronger the residents' purchasing ability which it is bound to the housing price. Similarly, the average disposable income (GDHI) determines the purchasing ability which is positively correlated with housing price (Mishkin., 2007).

Started and completed permanent dwellings (PDS) (PDC) and Consumer price inflation (CPI) are also chosen because of their importance in the housing market. Constructed area and completed dwellings numbers can best reflect the real estate supply. Usually in the case of constant elasticity of demand stability, as the supply area increases, the house price will decrease. Besides, CPI is introduced because it is normally considered as an indicator of Inflation or deflation which can lead to rising costs of workers or construction materials, indirectly affecting housing costs (Mishkin and SchmidtHebbel, 2001).

What's more, since the data of influencing factors of selected house prices are time-series data, there is a strong correlation between the data of the current year and the data of the previous year. Therefore, they should be pre-processed before making further analysis. In this paper, the increase of house price is predicted by the increase of each factor.

# 3 Methodology

## 3.1 Building initial regression formula

The relationship between variables should be checked in the beginning. The scatter plot is used and given below (Fig.2).

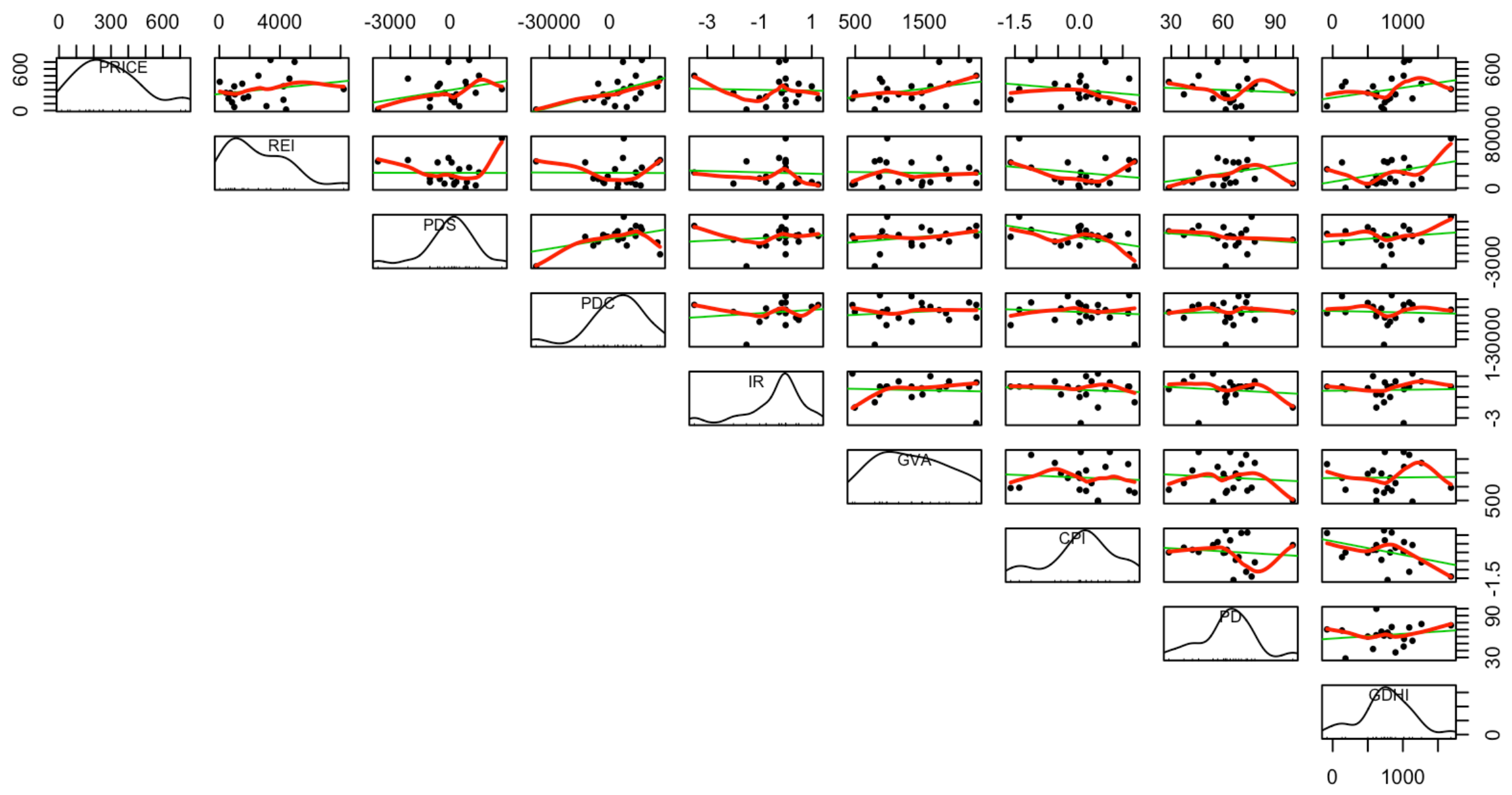

**Fig.2 Scatterplot matrix of dependent and independent variables**

As shown in figure2, each factor is inclined to a certain extent shape (linear), which indicates response variable and predictors have a linear relationship so that a linear regression model can be conducted. (RabeHesketh, 2008). Besides, there are no significant outliers.

The initial regression formula is:

$$IY = \beta_0 + \beta_1 REI + \beta_2 PDS + \beta_3 PDC + \beta_4 IR + \beta_5 GVA + \beta_6 CPI + \beta_7 PD + \beta_8 GDHI + \varepsilon$$

Among them, $\beta_0$ is the initial value of price changes; $\beta_1$ to $\beta_8$ are the regression coefficients and $\varepsilon$ means the residual.

## 3.2 Correlation Analysis to Avoid Multi-collinearity Problem

The correlation matrix below (Fig.3) represents the correlations between variables.

| | IY | REI | PDS | PDC | IR | GVA | CPI | PD | GDHI |
|---|---|---|---|---|---|---|---|---|---|
| IY | 1.00 | | | | | | | | |
| REI | 0.94 | 1.00 | | | | | | | |
| PDS | -0.43 | -0.56 | 1.00 | | | | | | |
| PDC | -0.18 | -0.26 | 0.76 | 1.00 | | | | | |
| IR | -0.88 | -0.84 | 0.64 | 0.55 | 1.00 | | | | |
| GVA | 0.99 | 0.93 | -0.50 | -0.28 | -0.91 | 1.00 | | | |
| CPI | 0.25 | 0.08 | -0.34 | -0.50 | -0.46 | 0.33 | 1.00 | | |
| PD | 0.98 | 0.95 | -0.57 | -0.30 | -0.91 | 0.99 | 0.30 | 1.00 | |
| GDHI | 0.99 | 0.94 | -0.53 | -0.30 | -0.92 | 0.99 | 0.30 | 1.00 | 1.00 |

**Fig.3 Correlation matrix between dependent variables and independent variables**

In figure 3, values in the first-row show that house price is strongly correlated with other independent variables, which provides the foundation for having a good linear relationship.

However, the high correlation values among predictor variables show a high level of multicollinearity and therefore it is unsuitable to direct input variables into the linear regression analysis. This problem can be overcome by entering the variables into Principal components analysis (PCA), as each PCA factor is uncorrelated (Dunteman, 1992). The details are introduced in the following parts.

## 3.3 Principal Components Analysis (PCA)

PCA is one of the most common dimension reduction technique. A great number of correlated variables are transformed into a limited set of uncorrelated variables which are called principal components (Dunteman, 1989).

### 3.3.1 Selecting the number of components to extract

Based on Kaiser–Harris criterion, components which eigenvalues greater than one should be retained (1964), so the Scree plot below indicates that there should be two principal components.

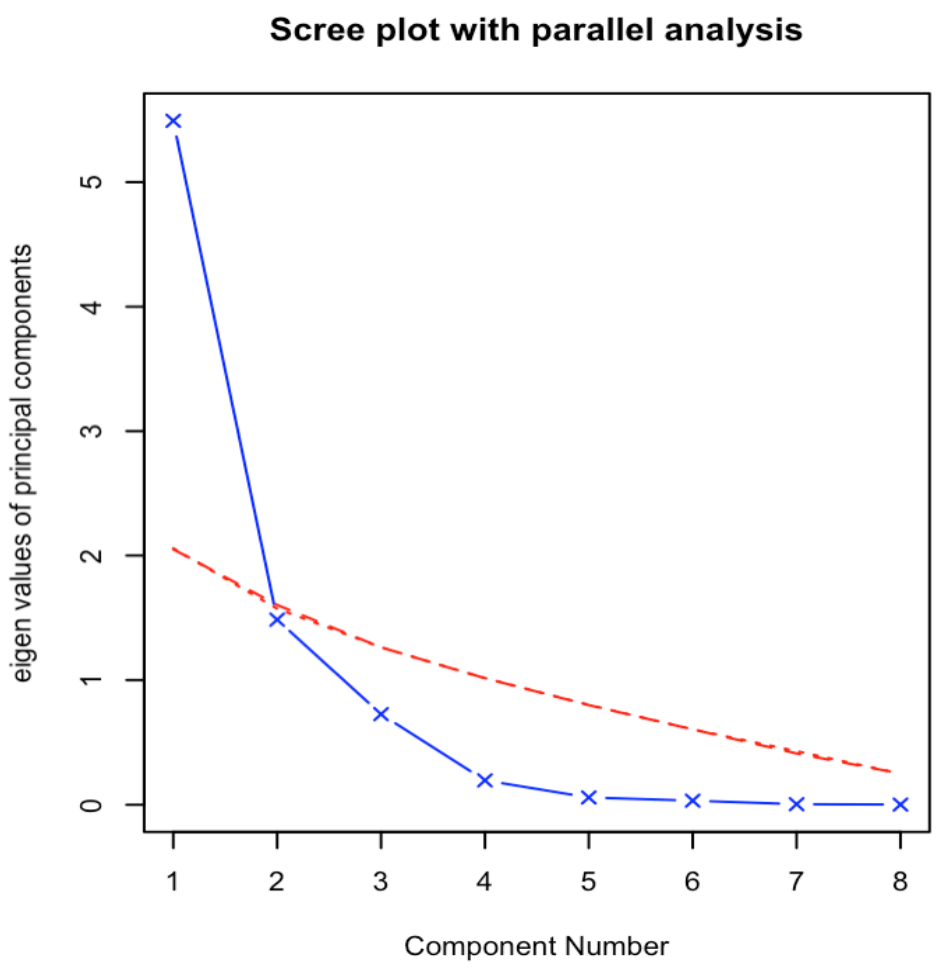

**Fig.4 Scree plots with eigenvalues**

### 3.3.2 Extracting and rotating principal components

After knowing the number of components, varimax rotations are needed to make the matrix with components more interpretable so that the number of variables can be reduced. The rotations results are provided in the following table (Kabacoff, 2010)..

| | RC1 | RC2 | h2 | u2 |
|---|---|---|---|---|
| REI | -0.1 | -0.97 | 0.95 | 0.0544 |
| PDS | -0.43 | 0.73 | 0.72 | 0.2785 |
| PDC | -0.13 | 0.92 | 0.86 | 0.139 |
| IR | -0.84 | 0.47 | 0.93 | 0.0673 |
| GVA | 0.97 | -0.19 | 0.98 | 0.0226 |
| CPI | 0.11 | -0.74 | 0.56 | 0.4364 |
| PD | 0.97 | -0.22 | 0.99 | 0.0091 |
| GDHI | 0.97 | -0.21 | 0.99 | 0.0149 |

| | RC1 | RC2 |
|---|---|---|
| SS loadings | 4.68 | 2.3 |
| Proportion Var | 0.59 | 0.29 |
| Cumulative Var | 0.59 | 0.87 |

**Fig.5 Components after rotating**

The correlations between variables and principal components can be observed in the column labeled RC1, RC2. The u2 column means the uniquenesses of components. For example, 97 percent of the variance in Gross value added (GVA) can be explained by the RC1. The further analysis can be conducted after obtaining the components scores.

Finally, the Proportion Var indicates that PCA1 account for the largest proportion of the variance (59%), while PCA2 makes up 29%. However, they are still different from principal components since the maximizing properties of the variance has not been preserved (Kabacoff, 2010).

### 3.3.3 Obtaining principal components scores

After extracting principal components and rotating principal components, we can obtain principal components scores using the formulas.

The principal components are:

$$PCA_1 = 0.23PD + 0.23GVA + 0.23GDHI + 0.23PDC + 0.14IR - 0.12CPI - 0.02PDS + 0.13REI$$

$$PCA_2 = 0.50PDC - 0.41CPI + 0.34PDS + 0.25REI + 0.10IR + 0.08GVA + 0.07PD + 0.08GDHI$$

The first component is strongly correlated with population density(PD), GVA and GDHI and appears to be a *demand* factor. Coefficients of them are all 0.23, which means they are equally vital to house price. Moreover, these variables vary together, so the increase of a variable will lead to the increase of other variables. Besides, due to the reason that the first principal component contributes the most to the model (Moore, 1981), it can be concluded that demands are now dominating and those three factors with highest weights are the most important factors to be taken into account.

The second component appears to be the *supply* factor, with high coefficients in started and completed permanent dwellings and real estate investment. The coefficient of started permanent dwellings is the highest (0.5), which means the number of newly constructed houses needs to be controlled. The weights of Interest rates(IR) are low, so it is not as important as other factors. CPI is negatively related to the principal component, which indicates that deflation may cause the more investment in the housing market.

The final linear regression model is as follows:

$$IY = \gamma_0 + \gamma_1 PCA_1 + \gamma_2 PCA_2 + \varepsilon$$

$\mathrm{IY}$ represents the annual increment of house prices; $\mathrm{PCA_1}$ and $\mathrm{PCA_2}$ are the principal components; $\varepsilon$ means the residual.

Then, the housing price model can be established

$$Y = Y_0 + IY$$

Here, $\mathrm{Y_0}$ represents the last year's housing price, $\mathrm{Y}$ represents the forecast of this year's house price.

# 4 Interpretation

| | Coefficients | Standard Error | t Stat | P-value |
|---|---|---|---|---|
| Intercept | -10340.99411 | 5927.041785 | -1.744714 | 0.104613 |
| REI | 0.017094598 | 0.012140529 | 1.4080604 | 0.182576 |
| PDS | 0.082354678 | 0.044281904 | 1.8597818 | 0.08569 |
| PDC | 0.004040739 | 0.003073729 | 1.314605 | 0.211364 |
| IR | -58.83223929 | 48.27708872 | -1.218637 | 0.244635 |
| GVA | 0.032579656 | 0.069879393 | 0.4662269 | 0.648769 |
| CPI | 10.07691633 | 66.06041967 | 0.1525409 | 0.881102 |
| PD | 1.865489384 | 1.627507066 | 1.1462251 | 0.272363 |
| GDHI | 0.074869309 | 0.090842246 | 0.8241684 | 0.424707 |

| Regression Statistics | |
|---|---|
| Multiple R | 0.997022 |
| R Square | 0.994053 |
| Adjusted R Square | 0.990394 |
| Significance F | 156.2394 |
| Observations | 22 |

**Figure 6. Statistical results of original multi-regression model**

The statistical results indicate that the original model is capable of explaining almost 100% of the variability in house price. However, the p-values indicate that the variables are not significant, so these variables are not a useful predictor within this model.

| Regression Statistics | |
|---|---|
| Multiple R | 0.98133527 |
| R Square | 0.96301892 |
| Adjusted R Square | 0.95912617 |
| Significance F | 322.286932 |
| Observations | 22 |

| | Coefficients | Standard Error | t Stat | P-value |
|---|---|---|---|---|
| Intercept | -1973.2416 | 522.8080875 | -3.7743134 | 0.00128281 |
| PCA1 | 0.27922603 | 0.01255322 | 22.2433789 | 4.5721E-15 |
| PCA2 | 0.06948691 | 0.006224727 | -11.163046 | 8.6969E-10 |

**Figure 7. Statistical results of adjusted multi-regression model**

The adjusted model can explain 96.3% of the variability in changes in real estate costs, which means a high degree of interpretation. Though the R-square decrease to a slight extent compared to the original one, its variables now all have significant p-values.

The housing price model can be re-written as:

$$Y = Y_0 + 0.279PCA_1 + 0.069PCA_2 - 1973.242$$

The coefficients indicate that those two components are positively correlated with real estate costs. To be more specific, for the explanation of the slope (coefficient) of model: if all else remain unchanged, for 100 units of PCA1 allocated, 27 housing price increment is explained. While on the slope of PCA2: all else held constant, for every 100 units allocated, 6.9 housing price increase is explained.

# 5 Limitation and discussion

The first limitation is that regarding dependent variable interpretation, PCA normally cannot be the theoretical explanation. They are the linear combination of influencing factors (DeCoster, 1998). So those principal components (PCA1, PCA2) are not effective enough to explain the housing price.

The other limitation of our model is that because we did not make predictions on influencing factors, this model can only calculate house price when knowing all independent variables. For further improvement, a predicting model of independent variables would be required and applying gray system theory may be a suitable solution (Kayacan et al., 2010).

## 6 Conclusion

This report analyzes the changes in London housing price from 1996 to 2016 and considers eight factors in terms of real estate industry supply and market demand. Through quantitative analysis of their potential relationship, a regression model of housing prices is built.

Population density, income, and GVA are the most significant factors that affect house price fluctuations in London which indicates that population and income are important point cuts to constrain housing increase. If the population and resident income are increased, housing demand will increase, causing the contradiction between supply and demand and, therefore, stimulating housing prices (Case, 2003).

Based on the above conclusions, the governments should develop the economy and reasonably guide the property demand. In the long term, the rational concept of housing consumption can restrain the real estate market bubble, so that the prices will return to rational level.